\documentclass[%
 reprint,
 amsmath,amssymb,
 superscriptaddress,
 aps,
 prl,
]{revtex4-2}

\usepackage{graphicx}    
\graphicspath{{Figures/finale/paper/}{Figures/finale/SM/}}
\usepackage[section]{placeins} 
\usepackage{float}             
\usepackage{dcolumn}     
\usepackage{hyperref}  

\usepackage{bm}          
\usepackage{amsmath,amsfonts,amssymb}
\usepackage{siunitx}
\usepackage{orcidlink}

\DeclareMathOperator{\Imag}{Im}

\begin{document}

\preprint{APS/123-QED}

\title{Graphs on chip: a silicon photonics platform}

\author{H. Girin, \orcidlink{0009-0000-5449-9917}}
\email{hugo.girin@cnrs.fr}
\affiliation{Université Paris-Saclay, CNRS, Centre de Nanosciences et de Nanotechnologies, 91120 Palaiseau, France.}

\author{X. Checoury}
\email{xavier.checoury@cnrs.fr}
\affiliation{Université Paris-Saclay, CNRS, Centre de Nanosciences et de Nanotechnologies, 91120 Palaiseau, France.}

\author{B. Odouard}
\affiliation{Université Paris-Saclay, CNRS, Centre de Nanosciences et de Nanotechnologies, 91120 Palaiseau, France.}

\author{S. Bittner, \orcidlink{0000-0003-3651-5762}}
\affiliation{Université de Lorraine, CentraleSupélec, LMOPS EA-4423, 57070 Metz, France.}
\affiliation{Chaire Photonique, LMOPS EA-4423, Centralesup\'elec, 57070 Metz, France}

\author{J.-R.\ Coudevylle}
\affiliation{Université Paris-Saclay, CNRS, Centre de Nanosciences et de Nanotechnologies, 91120 Palaiseau, France.}

\author{B. Dietz,\orcidlink{0000-0002-8251-6531}}
\email{bdietzp@pks.mpg.de }
\affiliation{Max-Planck Institute for the Physics of Complex Systems, N\"othnitzer Stra\ss e 38, 01187 Dresden, Germany.}
\affiliation{TU Dresden, Institute of Theoretical Physics, 01062 Dresden, Germany.}

\author{C. Lafargue}
\affiliation{Laboratoire Lumière, Matière et Interfaces (LuMIn),
CNRS, ENS Paris-Saclay, Université Paris-Saclay, CentraleSupélec,
91190 Gif-sur-Yvette, France.}

\author{M. Lebental}
\email{melanie.lebental@cnrs.fr}
\affiliation{Université Paris-Saclay, CNRS, Centre de Nanosciences et de Nanotechnologies, 91120 Palaiseau, France.}
\affiliation{ENS Paris-Saclay, 91190  Gif-sur-Yvette, France.}

\date{\today} 


\begin{abstract}
We present a versatile photonic waveguide network platform implementing quantum (wave) graphs, proposed by Kottos \& Smilansky [Phys. Rev. Lett. \textbf{79}, 4794 (1997)] to investigate ray-wave correspondence and quantum chaos.
Realized on a silicon-on-insulator chip at telecom wavelengths, it enables a unique capability: the direct, non-invasive imaging of individual optical wave functions with unprecedented resolution via third-harmonic generation in silicon, which we use to study the localization of intensity distributions.
Furthermore, by investigating two graphs with contrasting classical dynamics --- one strongly chaotic, the other one ergodic, but non-mixing --- we reveal strikingly different spectral statistics, in quantitative agreement with random matrix theory and a minimal unitary quantum-map model.
This establishes silicon photonics networks as a versatile and scalable platform for investigating quantum chaos and non-linear graphs, and paves the way for optical quantum computing in complex networks.
\end{abstract}

\maketitle

\begin{figure}[ht]
    \centering
    \includegraphics[width=1\linewidth]{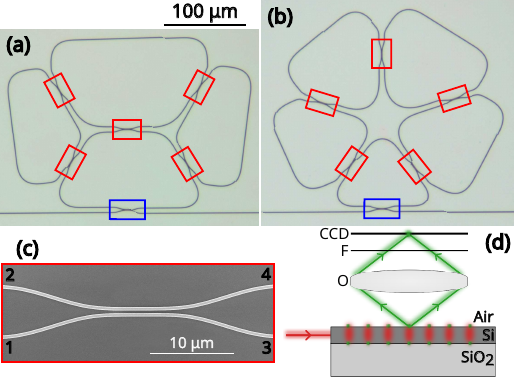}
    \caption{
        Optical images of (a) a BTG and (b) a FG.
        Red boxes highlight the 2:2 bidirectional couplers forming the internal connections of the graph,
        while blue boxes mark the couplers used to inject and collect light.
        (c) Scanning Electron Microscope (SEM) image of a 2:2 bidirectional coupler.
        (d) Layout of the setup for intensity distribution imaging.
        Infrared light guided at \(\lambda \approx 1550\)~nm generates a THG signal at \(\lambda/3 \approx 515\)~nm that is not guided and radiates out of the waveguide plane. It is collected by an objective (O), filtered (F) to suppress residual second-harmonic background, and imaged on a visible (EM)CCD camera (Andor iXon Ultra 897).
    }
    \label{fig:exp_graphs}
\end{figure}

Graphs — mathematical structures of vertices connected by edges~\cite{diestel2017} — provide a ubiquitous and analytically tractable framework for modeling complex systems across mathematics and physics.
They capture the essential connectivity of complex systems in a compact and computable form, with applications ranging from combinatorics~\cite{bollobas1998} and information theory~\cite{kschischang2001} to condensed-matter physics~\cite{anderson1958,yin2022} and quantum computing~\cite{raussendorf2001,thomas2024}.
A rich instantiation is the \emph{wave graph} (or quantum graph)~\cite{kottos1997,kottos1999,kuchment2003,berkolaiko2012,berkolaiko2026}: a network of scattering vertices, where wave propagation on each bond follows a one-dimensionnal Schrödinger equation~\cite{gnutzmann2006}.
Despite their apparent simplicity, wave graphs display complex spectral and transport properties governed by the network topology, and serve as controllable models for wave chaos, localization~\cite{klopp2009}, and ergodicity.
Hence, they have become paradigmatic models for studying quantum chaos~\cite{bohigas1984}, with spectral statistics predicted to follow random matrix theory when the classical dynamics is mixing~\cite{pluhar2013,pluhar2014}.

Several experimental platforms have been developed to realize wave graphs, like fiber networks \cite{lepri2017}, and most prominently microwave cable networks~\cite{hul2004} constructed
from coaxial cables connected by T-joints, with which all random-matrix universality classes have been implemented~\cite{dyson1962,altland1997,lawniczak2010,bialous2016,
rehemanjiang2016,martinez-arguello2018,rehemanjiang2020,lu2020,che2021}. While these approaches provide excellent controllability, they suffer from backscattering at junctions, which allow
non-universal modes~\cite{dietz2017} localized on a small fraction of the graph. Furthermore, wave functions can be probed only at the vertices~\cite{exner2015} for cable networks, or via a perturbation body~\cite{zhang2022} for waveguide graphs.

In this Letter, we introduce silicon on-chip waveguide networks as a powerful experimental platform to study wave graphs. Integrated photonic networks \cite{ma2022,wang2023} have already been explored for diverse applications, including high-finesse spectroscopy~\cite{zhang2021}, random laser networks~\cite{saxena2025}, and optical quantum computing~\cite{ehrhardt2021,qiang2021,bao2023}. Our wave graph vertices are implemented by \emph{2:2 bidirectional couplers},
which consist of two parallel waveguides coupled by evanescent tunneling [Fig.~\ref{fig:exp_graphs}(c)].
Such couplers eliminate backscattering and ensure equal redistribution of optical power.
Our devices operate at the telecom wavelength $\lambda_0 = 1.55~\mu$m and offer several additional advantages:
they function at room temperature, are scalable and compact due to the short optical wavelength, are compatible with existing telecom technology,
and fully integrated on chip.
Most importantly, they allow \emph{direct imaging of the intensity distributions} via third-harmonic generation (THG) in silicon~\cite{zeng2016} [Fig.~\ref{fig:exp_graphs}(d)],
providing access to spatial information with unprecedented accuracy.

The main goal of this Letter is to test whether our silicon-integrated wave graphs are suitable for studying quantum chaos.
To this end, we focus on two representative five-vertex networks: a \emph{Flower graph} (FG), which is ergodic but non-mixing, and a \emph{Bow-Tie graph} (BTG), which exhibits mixing dynamics.
We find that the spectral statistics of the BTG is consistent with Gaussian Orthogonal Ensemble (GOE) statistics, while the FG shows clear deviations from universality.
These results establish silicon photonics waveguide networks as a most suitable system for probing how graph topology governs the emergence of complexity and wave-dynamical chaos.

The Letter is organized as follows.
First, we present the silicon photonics waveguide network and the setup for optical characterization.
Then the theoretical framework of wave graphs is introduced and compared with the experimental spectral statistics of the FG and BTG, with a focus on the nearest-neighbor spacing distribution and a variant of the length spectrum. Finally we demonstrate direct imaging of intensity distributions using THG.\\

\textbf{Experiments.} The network consists of rib waveguides interconnected at bidirectional couplers to form closed resonant structures [Fig.~\ref{fig:exp_graphs}(a-b)]. The lengths of the individual bonds are chosen to be \emph{incommensurate} (Supplemental Material, Sec.~\ref{app:graph_topo}), thereby avoiding systematic degeneracies in the resonance spectrum.
The inclusion of a bus waveguide for optical injection and collection renders the network an open system.

The devices were fabricated on a silicon-on-insulator (SOI) platform using standard electron-beam lithography and dry-etching processes, see End Matter for details. The resulting rib waveguides ensure single-mode operation for the fundamental transverse-electric (TE) mode~\cite{jackson1999} at 1.5~µm wavelength, with an effective index $n_{\textrm{eff}}\simeq 2.4$. The coupling between adjacent waveguides is achieved through 2:2 bidirectional couplers [cf.~Fig.~\ref{fig:exp_graphs}(c)].
The coupler parameters were optimized following Ref.~\cite{gupta2017} to achieve a 50:50 power splitting ratio around 1.5~$\mu$m while minimizing wavelength dependence, see End Matter.\\

\begin{figure}[ht]
    \centering
    \includegraphics[width=0.9\linewidth]{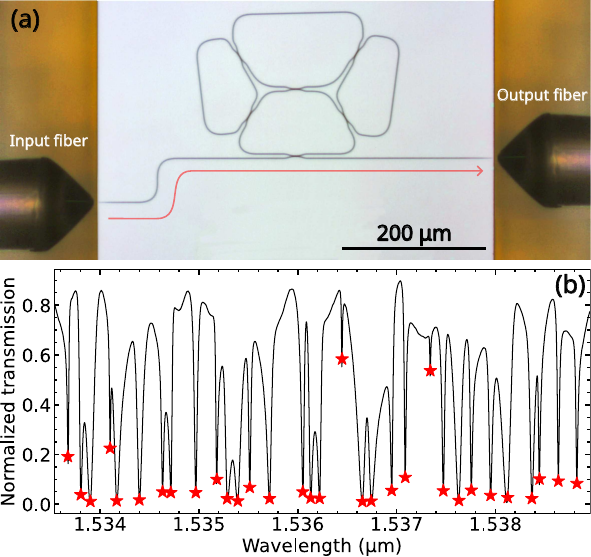}
    \caption{(a) Photograph of the experimental setup. An S-shaped bus waveguide (red arrow) is used to prevent any direct parasitic coupling between the fibers that could bypass the photonic network.
        (b) Zoom of the normalized transmission spectrum of the BTG.
        The dips (red stars) correspond to optical resonances of the network. The transmission is normalized by the propagation through an identical S-shaped bus waveguide without the network.
    }
    \label{fig:exp_setup}
\end{figure}

The optical characterization setup is shown in Fig.~\ref{fig:exp_setup}(a).
A tunable laser (Santec TSL-710) sweeping from 1480~nm to 1640~nm provides continuous-wave TE-polarized light that is injected into the input waveguide using a lensed fiber. The transmitted light is collected by another lensed fiber and recorded by a photodetector.

A typical transmission spectrum measured for the BTG is shown in Fig.~\ref{fig:exp_setup}(b).
It exhibits well-resolved resonance dips with quality factors reaching up to \(Q \approx 2 \times 10^5\), and a finesse of approximately $F \approx 5$ \footnote{In this context, the finesse $F$ is defined as the mean resonance spacing divided by the mean linewidth.}, which provides sufficient spectral resolution to accurately analyze the resonance structure of the network.\\

\textbf{Wave Model.} We now introduce the theoretical framework used to model our waveguide networks.
The formalism originates from the theory of wave graphs. Although initially developed in the context of quantum mechanics, this approach applies equally well to classical wave systems.

A waveguide network is composed of $V$ vertices (couplers) connected by $B$ bonds (waveguides), each bond $b$ having a length $L_b$.
Our silicon waveguides support only the fundamental TE mode at telecom wavelengths; hence, the optical field on each bond can be described by a one-dimensional scalar wave function. On a given bond $b$, the field is written as a superposition of counter-propagating plane waves,
\begin{equation}\label{eq:ondes-planes}
    \psi_b(x)
    = a_b^{+}\, e^{\, i n_\mathrm{eff} k x}
    + a_b^{-}\, e^{- i n_\mathrm{eff} k x},
\end{equation}
where $x \in [0,L_b]$ is the coordinate along the bond, $k=2\pi/\lambda$ is the vacuum wavenumber, and $n_\mathrm{eff}$ is the effective index of the guided mode.
The complex coefficients $a_{b}^{+}$ and $a_{b}^{-}$ depend on the boundary conditions imposed at the vertices, namely continuity of the wave function, and current conservation. All these conditions result in a linear system, that can be written as \cite{wang2023}
\begin{equation}\label{eq:U-operator}
\mathcal{U}_B(k) \textbf{A}=\textbf{A}
\end{equation}
where $\textbf{A}$ is a vector gathering the 2$B$ coefficients $a_b^{\pm}$ and $\mathcal{U}_B(k)$ is a 2$B\times 2B$ matrix, also called the global evolution operator. The network spectrum is given by the real solutions $\{k_p\}$ of Eq.~\ref{eq:U-operator}, and the corresponding $\{\textbf{A}(k_p)\}$ are the eigenfunctions.

\begin{figure}[ht]
    \centering
    \includegraphics[width=1.0\linewidth]{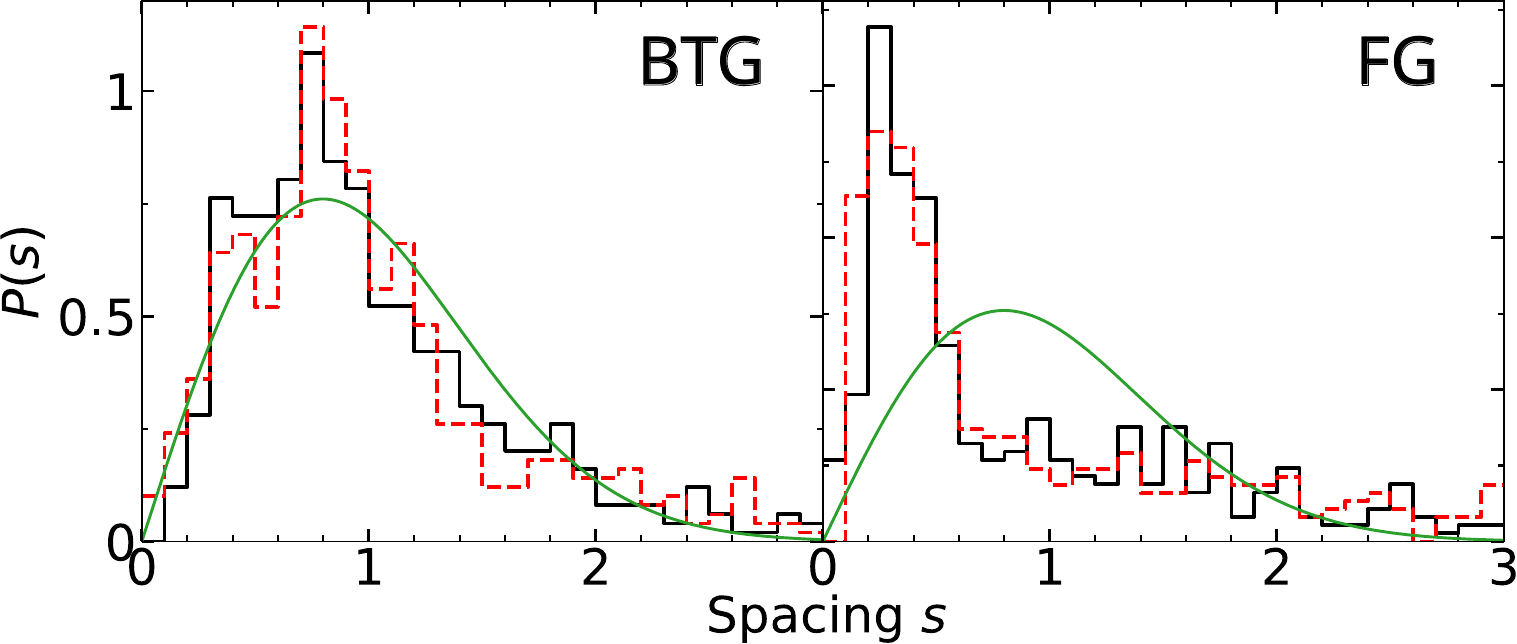}
    \caption{Nearest-neighbor spacing distribution $P(s)$ of resonances (see Sec.~\ref{app:unfolding} of the Supplemental Material) for
    (a) BTG and (b) FG. The black solid histogram shows the experimental
    data (normalized to unit area), the red dashed curve the results for the closed-graph model, and the green solid curve the Wigner distribution of the GOE.
    Panel (a) is obtained from 502 resonances measured in the wavelength range $\lambda \in [1.48,\,1.57]$~µm.
    Panel (b) corresponds to 394 resonances in $\lambda \in [1.48,\,1.55]$~µm. In both cases, the effective
    refractive index in the theoretical model was adjusted such that the predicted number of modes matches
    the experimental number within the same spectral window. The fraction of missing resonances
    is estimated to be below $2.5~\%$ for both devices.}
    \label{fig:NNSD}
\end{figure}

This closed-graph model is sufficient to describe the experimental spectral statistics reported in Fig.~\ref{fig:NNSD}.
We can also calculate the transmission spectrum $|\mathcal{T}(k)|^2$ by including the coupling between the graph and the bus waveguide; this is treated using the open-graph formalism described in Sec.~\ref{app:open_graphs} of the Supplemental Material.

As shown in \cite{wang2023}, the linear operator $\mathcal{U}_B$ is actually a product of two matrices $\mathcal{U}_B=P(k)\,\Sigma$. The matrix $P(k)$ includes all the phase accumulations $e^{\pm i n_\mathrm{eff} k L_b}$ and thus depends on $k$, while the matrix $\Sigma=(\Sigma_{b',b})$ is composed of the local scattering matrices at the vertices $\{\sigma\}_V$, and is considered to be independent of $k$. The End Matter gives more details on $\sigma$ for our 2:2 bidirectional coupler.\\

\textbf{Classical limit: ergodicity and mixing.} The classical (ray) limit of the quantum map $\mathcal{U}_B$ governs the evolution of a particle on the graph and defines a discrete-time Markov process, where the transition probability from bond $b$ to bond $b'$ reads~\cite{gnutzmann2006}
\begin{equation}\label{eq:def-Perron-Frobenus}
    \mathcal{F}_{b',b} = |\Sigma_{b',b}|^2.
\end{equation}
Defining a vector $\textbf{V}_0$ of dimension $2\mathcal{B}$, whose components equal the initial occupation probabilities on the $2\mathcal{B}$ directed bonds, their values after $m$ time steps are given as
\begin{equation}\label{eq:Markov-classique}
\textbf{V}_m=\mathcal{F}^m\textbf{V}_0
\end{equation}
with $\mathcal{F}$ denoting the Perron-Frobenius operator~\cite{gnutzmann2006}.
Since $\sigma$ is lossless, which is indeed the case experimentally, $\mathcal{F}$ is bistochastic~\footnote{The elements $a_{ij}$ of a bistochastic matrix $A$ are confined to $1\geq a_{ij}\geq 0$ and their sums over each row and each column equal unity, $\sum_i a_{ij}=\sum_j a_{ij}=1$.} and the uniform distribution with equal occupation probabilities $1/(2\mathcal{B})$ on all bonds is invariant. Hence, its spectrum, which is restricted to the unit circle and its interior, always comprises one eigenvalue equal to unity with all components of the corresponding eigenvector equal to $1/(2\mathcal{B})$. A dynamically connected graph has an \emph{ergodic} classical dynamics~\footnote{A graph is dynamically connected if it cannot be split into subsets such that probabilities for the transition between any of the sets is zero. This also excludes Dirichlet boundary conditions at a vertex.}~\cite{kottos1999,gnutzmann2006}. A stronger type of chaos, called \emph{mixing}, appears if $1$ is the only eigenvalue of $\mathcal{F}$ with unit modulus, implying that any initial probability distribution $\textbf{V}_0$ tends to the uniform distribution with increasing $m\to\infty$. As shown in Sec.~\ref{app:PF} of the Supplemental Material, the BTG is mixing, whereas the FG is ergodic but not mixing.\\

\textbf{Spectral Statistics.} According to the Bohigas-Giannoni-Schmit (BGS) conjecture, the spectral fluctuations of the mixing BTG are expected to follow GOE statistics, up to finite size effects and cycles, as evidenced in Refs.~\cite{lepri2017,gnutzmann2013}. Short-range spectral correlations are characterized in Fig.~\ref{fig:NNSD} through the nearest-neighbor spacing distribution (NNSD), computed as detailed in Sec.~\ref{app:unfolding} of the Supplemental Material.

For BTG and FG, the experimental data agree quantitatively with the
closed-graph model, demonstrating that the simple unitary
quantum-map description captures the observed spectral fluctuations without explicitly
including the bus waveguide, dispersion, or coupler wavelength dependence.

A significant difference, however, appears between the two topologies.
For the mixing BTG, the spacing distribution approaches the GOE curve, exhibiting clear level repulsion.
Residual deviations can be attributed to finite-size effects and cycles, known to induce non-universal corrections in wave graphs~\cite{lawniczak2010,dietz2017,gnutzmann2013}. It agrees well with the Rosenzweig-Porter distribution for $\lambda \simeq 0.5$ (not shown) \cite{cadez2024}.
By contrast, the FG displays pronounced deviations from GOE statistics, reflecting the absence of full classical mixing.

Additional long-range correlations, namely the number variance $\Sigma^2$ and the spectral rigidity $\Delta_3$, are reported in Sec.~\ref{app:additional_stats} of the Supplemental Material and confirm this behavior.\\

\begin{figure}[t]
    \centering
    \includegraphics[width=1.0\linewidth]{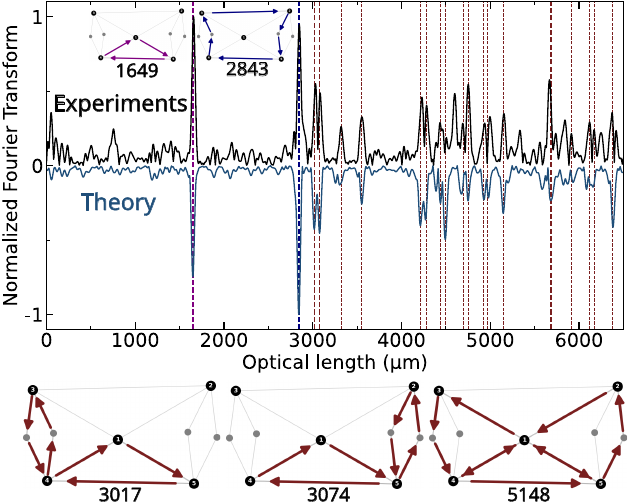}
    \caption{
    Fourier transform of the transmission spectrum $|\mathcal{T}(k)|^2$ of the BTG over the wavelength range $1480$--$1600$~nm.
    The experimental spectrum (top) and that for the open-graph model (bottom) are independently normalized to unit maximum amplitude.
    Vertical dashed lines mark the optical lengths $L_{\mathrm{PO}} = n_{\mathrm{g}}\,\ell_{\mathrm{PO}}$ of selected periodic orbits, where the group index $n_{\mathrm{g}}=3.97$ is obtained from simulation for the fundamental TE mode at $1.55~\mu$m.
    Two periodic orbits are exhibited close to their corresponding spectral peaks.
    Bottom panel: diagrams of three additional periodic orbits with their optical lengths (in~$\mu$m).}
    \label{fig:length_spectrum}
\end{figure}

\textbf{Manifestation of periodic orbits.} Insight into the periodic orbits of the classical dynamics is obtained for the BTG from the Fourier transform of the experimental transmission $|\mathcal{T}(k)|^2$ and compared with the open-graph model including coupling to the bus waveguide (Sec.~\ref{app:open_graphs} of the Supplemental Material). The agreement between experiment and theory is excellent.

Like for the Fourier transform of the spectral density or of its semiclassical approximation in terms of a trace formula, called length spectrum~\cite{gnutzmann2006}, the positions of the dominant peaks correspond to lengths of periodic orbits of the graph.
The two strongest peaks around 1600~$\mu$m and 3000 $\mu$m arise from two loops directly coupled to the bus waveguide, providing the shortest and most efficiently excited trajectories. Their presence partly explains deviations of the spectral properties of the BTG from GOE statistics. Additional weaker peaks are associated with more complex periodic orbits involving multiple bonds.

These results demonstrate that the interference structure of the transmission is governed by the classical dynamics of the underlying graph both for the FG (not shown) and the BTG.\\

\begin{figure*}
    \includegraphics[width=1.0\textwidth]{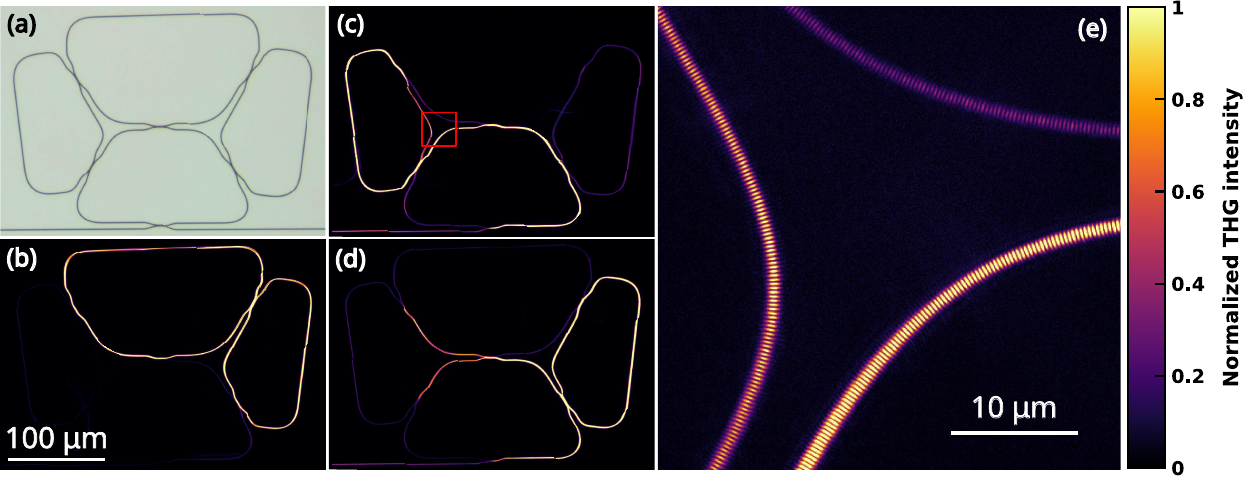}
    \caption{ (a) Optical image of the BTG under white-light illumination.
    (b-d) Near-field images obtained with the setup in Fig.~\ref{fig:exp_graphs}(d). The excitation wavelengths are (b) 1528.4 nm,
    (c) 1530.5 nm, and (d) 1531.4 nm. Each image is displayed on a linear intensity scale and normalized such that the maximum pixel intensity reaches the full dynamic range. The full device image is reconstructed by merging six partially overlapping frames.
    (e) Magnified view of the region indicated by the red square in (c).}
    \label{fig:exp_wavefunction}
\end{figure*}

\textbf{Intensity-Distribution Imaging.} A crucial advantage of our silicon platform is the possibility to directly image the optical intensity distribution inside the graph via third-harmonic generation: continuous-wave excitation around 1550~nm produces visible emission near 515~nm, which is not guided by the waveguides and radiates out of plane [Fig.~\ref{fig:exp_graphs}(d)]. Note that an efficient collection of the radiated THG requires a numerical aperture larger than $n_\mathrm{eff}/3 \simeq 0.8$.
An erbium-doped fiber amplifier boosts the injected power sufficiently to generate a detectable THG signal via resonant build-up inside the graph (Sec.~\ref{app:power_budget} of the Supplemental Material).

Because the THG wavelength is shorter than the infrared excitation wavelength in the material ($\lambda_\mathrm{IR}/n_\mathrm{eff}\approx 640$~nm), this approach enables sub-wavelength spatial resolution while remaining minimally invasive.

Figure~\ref{fig:exp_wavefunction} shows three experimental resonant modes of the BTG at nearby wavelengths.
The spatial intensity distributions differ markedly reflecting the mode-to-mode fluctuations expected in complex wave systems.
These measurements provide access to spatial observables that are typically inaccessible in other experimental realizations of wave graphs~\cite{hul2009, kaplan2001, zhang2022}.
In total, we measured 4 intensity profiles on the BTG, of which 3 are displayed in Fig.~\ref{fig:exp_wavefunction}.
They enable quantitative analysis of wave function localization. We extract a normalized entropy quantifier (Sec.~\ref{app:localization} of the Supplemental Material), which is equal to 0 for complete localization on a single bond and 1 for perfectly uniform delocalization.
Averaged over all measured modes, we obtain $0.93 \pm 0.02$ experimentally in the BTG graph, in good agreement with $0.90 \pm 0.04$, which is inferred from the eigenfunctions of the closed-graph model. These results are consistent with the tendency towards quantum ergodicity in mixing graphs~\cite{anantharaman2019}. A systematic statistical study of localization will be presented elsewhere.

The magnified view in Fig.~\ref{fig:exp_wavefunction}(e) resolves the spatial modulation of the standing wave along a single bond. The effective index can be extracted from the measured fringe spacing. At $\lambda = 1530.5~\mathrm{nm}$, we obtain $n_{\mathrm{eff}} = 2.410 \pm 0.014$, in excellent agreement with simulations (Sec.~\ref{app:measure_neff} of the Supplemental Material).

These results establish nonlinear imaging as a powerful and quantitative probe of wave dynamics in integrated wave graphs, providing direct access not only to spectral statistics, but also to the spatial structure of individual eigenmodes.\\

\textbf{Conclusion.}  The excellent agreement between our experimental platform and the minimal quantum-map model confirms that silicon photonics waveguide networks reliably implement the physics of wave graphs. By comparing two five-vertex graphs with distinct classical dynamics, we confirm a direct link between mixing properties and spectral universality: the Bow-Tie graph roughly follows GOE statistics, whereas the non-mixing Flower graph exhibits clear deviations. A central feature of our platform is the direct imaging of optical intensity distributions via third-harmonic generation, which enables systematic studies of wave function localization in complex graphs.

Operating at room temperature, fully compatible with CMOS fabrication, and scalable to large networks at telecom wavelengths, silicon photonics networks offer a versatile platform to explore nonlinear wave dynamics in complex systems and pave the way to optical quantum computing.\\

\begin{acknowledgments}
This work was done with the C2N micro- and nano-technology platform. It was supported by the French RENATECH network, and by the departement de l'Essonne. BD is partially funded by the Deutsche Forschungsgemeinschaft (DFG, German Research Foundation) Project No.~290128388. S.B.\ acknowledges support for the Chaire Photonique by R\'egion Grand Est, GDI Simulation, D\'epartement de la Moselle, European Regional Development Fund, CentraleSup\'elec, Fondation CentraleSup\'elec and the Eurometropole de Metz. \\
C. Eggenspieler, S. Nonnenmacher and J. Zyss are acknowledged for fruitful discussions, E. Herth and D. Bouville for technical support, and J. Erb for sharing some of his experimental data.
\end{acknowledgments}


\newpage 
\appendix

\begin{center}
 \Large\textbf{End Matter}\end{center}
 \normalsize
\section{Appendix: Waveguide design and fabrication}
\label{app:fabrication}

\textbf{Fabrication process.} The devices were fabricated on a silicon-on-insulator (SOI) substrate consisting of a 220~nm-thick silicon layer on a 2~µm-thick silica (SiO$_2$) buried oxide layer. Patterns were defined by electron-beam lithography (EBPG5200) using hydrogen silsesquioxane (HSQ) resist, followed by inductively coupled plasma reactive ion etching (ICP-RIE) with an SPTS system using SF$_6$ and C$_4$F$_8$ gases. The remaining HSQ resist is removed by hydrofluoric acid (HF). The resulting rib waveguides have a partial etch geometry described below.\\

\begin{figure}[ht]
    \centering
    \includegraphics[width=1.0\linewidth]{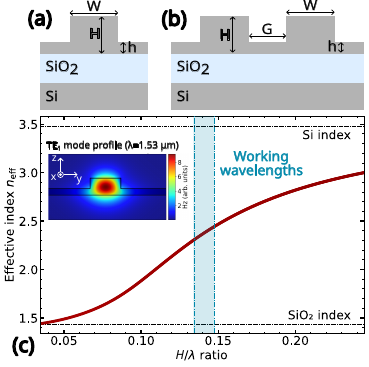}
    \caption{
        (a)~Schematic cross-sectional view of a single rib waveguide with total silicon height $H = 220$~nm, ridge width $W = 400$~nm, and slab thickness $h = 85$~nm. Not to scale.
        (b)~Schematic cross-sectional view of the directional coupler, sharing the same $W$, $H$, and $h$ as the single waveguide, with gap $G = 300$~nm between the two ridges. Not to scale.
        (c)~Effective index $n_\mathrm{eff}$ of the fundamental quasi-TE mode as a function of $H/\lambda$ ($H=220$ nm), computed with MPB. The shaded blue region indicates the working wavelength range ($\lambda = 1.48$--$1.64$~µm). Inset: $H_z$ field profile of the fundamental quasi-TE mode at $\lambda = 1.53$~µm.
    }
    \label{fig:neff_MPB}
\end{figure}

\textbf{Waveguide geometry and mode properties.}
The rib waveguides are characterized by a total silicon height $H = 220$~nm, a ridge width $W = 400$~nm, and a slab thickness $h = 85$~nm [see Fig.~\ref{fig:neff_MPB}(a)]. The slab thickness $h$ was selected as the key design parameter to balance two competing requirements: minimizing bend losses (by means of a strongly confining, deeply etched geometry) and reducing the spectral dependence of the directional coupler response (by means of a shallower etch). Bends are implemented as Euler curves with a minimum radius of 30~µm, which reduces mode-mismatch losses at the transitions.

To ensure efficient and broadband fiber-to-chip coupling, the input and output waveguides are terminated by adiabatic inverse tapers that linearly reduce the ridge width from 400~nm to 150~nm over 40~µm, expanding the optical mode to better match the field profile of the lensed fiber (spot diameter ${\approx}\,2$~µm).

Because the waveguide cladding is asymmetric --- silica below and air above --- the guided modes lack vertical mirror symmetry and are therefore not purely transverse-electric. Full-vectorial simulations using MPB (MIT Photonic Bands~\cite{johnson2001}) show that the fraction of optical power carried by the TE field components ranges from 96.7\% at $\lambda = 1.64$~µm to 96.8\% at $\lambda = 1.48$~µm. The waveguides thus support a quasi-TE fundamental mode and are single-mode over the entire measurement bandwidth.

\textbf{Effective index.} The effective index $n_\mathrm{eff}$ of the fundamental quasi-TE mode was computed with MPB as a function of $H/\lambda$, where $H = 220$~nm is the total silicon thickness and $\lambda$ is the free-space wavelength [Fig.~\ref{fig:neff_MPB}(c)]. At the design wavelength $\lambda = 1550$~nm ($H/\lambda = 0.142$), the simulation gives $n_\mathrm{eff} = 2.38(9)$. At $\lambda = 1531$~nm, the wavelength used for experimental effective-index measurements via intensity distribution imaging (Sec.~\ref{app:measure_neff} of the Supplemental Material), we obtain $n_\mathrm{eff} = 2.40(5)$.

\section{Appendix: 2:2 bidirectional couplers}
\label{app:coupler_design}

Bidirectional couplers constitute the elementary scattering vertices of the photonic graphs investigated in this Letter.
Their design critically determines the effective vertex scattering matrix $\sigma$ and therefore the global mixing properties of the network.\\

\begin{figure}[H]
    \vspace{6pt}
    \centering
    \includegraphics[width=0.9\linewidth]{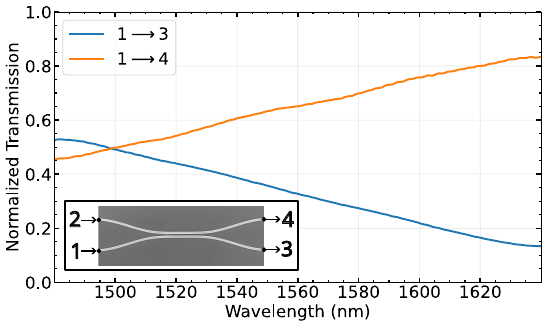}
    \caption{
    Experimental normalized transmission of the bidirectional coupler from port 1 to ports 3 (bar) and 4 (cross).
    The transmission is normalized by a reference bus waveguide without network.
    Inset: SEM image of the device including port numbering.}
    \label{fig:coupler_design_exp}
\end{figure}

\textbf{Theoretical background.} For our 4-channel coupler, $\sigma$ is a 4$\times$4 matrix. It is symmetric for a bidirectional coupler, and furthermore unitary (i.e. $\sigma^{\dagger}\sigma=I_d$) if it is lossless. It reads
\begin{displaymath}
\sigma=
\left(\begin{array}{cccc} 0&0&\sqrt{1-C} & i\sqrt{C}\\
0&0&i\sqrt{C} & \sqrt{1-C}\\
\sqrt{1-C} & i\sqrt{C}&0&0\\
i\sqrt{C} & \sqrt{1-C}&0&0
\end{array}\right)
\end{displaymath}
where $C$ is the coupling parameter, and is equal to 0.5 for a well-balanced coupler. The factor $i$ arises from the relative $\pi/2$ phase shift between the transmitted (bar) and coupled (cross) amplitudes, which results from the interference between the symmetric and antisymmetric supermodes of the directional coupler~\cite{berkolaiko2012,ansys_coupler}. For the closed-graph simulations presented in Fig.~\ref{fig:NNSD}, we used $C=0.5$, which is thus independent of $k$. For the open graph model employed in Fig.~\ref{fig:length_spectrum}, the coupling parameter $C$ is inferred from the experimental coupler shown in Fig.~\ref{fig:coupler_design_exp}, and depends then on the wavelength~\cite{berkolaiko2012}.\\

\textbf{Coupled propagating waves.} In a lossless bidirectional coupler supporting two coupled TE supermodes, the transmitted powers at the bar and cross ports (cf.~Fig.~\ref{fig:exp_graphs}(c)) can be derived from coupled-mode theory~\cite{gupta2017} as
\begin{align}
P_{1\rightarrow 3}(\lambda) &=
\cos^2\!\left[\frac{\pi \, \Delta n_{\mathrm{eff}}(\lambda)}{\lambda} \, l_{\mathrm{DC}}\right] P_{\mathrm{in}},
\\
P_{1\rightarrow 4}(\lambda) &=
\sin^2\!\left[\frac{\pi \, \Delta n_{\mathrm{eff}}(\lambda)}{\lambda} \, l_{\mathrm{DC}}\right] P_{\mathrm{in}},
\end{align}
where $\Delta n(\lambda)=n_{\mathrm{eff}}^{(S)}-n_{\mathrm{eff}}^{(AS)}$ is the effective index difference between the symmetric and antisymmetric supermodes, $l_{\mathrm{DC}}$ is the coupling length, and $P_{\mathrm{in}}$ is the wavelength independent input power.

The coupling length, which yields a 50:50 splitting ratio, is given by
\begin{equation}
l_{50} = \frac{\lambda}{4\,\Delta n_{\mathrm{eff}}}.
\end{equation}

The spectral dependence of the splitting ratio is governed by the differential group index
\begin{equation}
\Delta n_g = \Delta n_{\mathrm{eff}}(\lambda)
- \lambda \frac{d}{d\lambda}\big[\Delta n_{\mathrm{eff}}(\lambda)\big].
\end{equation}
When $\Delta n_g \rightarrow 0$, the splitting ratio becomes nearly wavelength independent over a finite spectral range.
This design strategy, proposed in Ref.~\cite{gupta2017}, consists in jointly tuning the waveguide width $W$, slab height $h$, and gap $G$ in order to minimize $\Delta n_g$.

\begin{figure}[ht]
    \centering
    \includegraphics[width=0.9\linewidth]{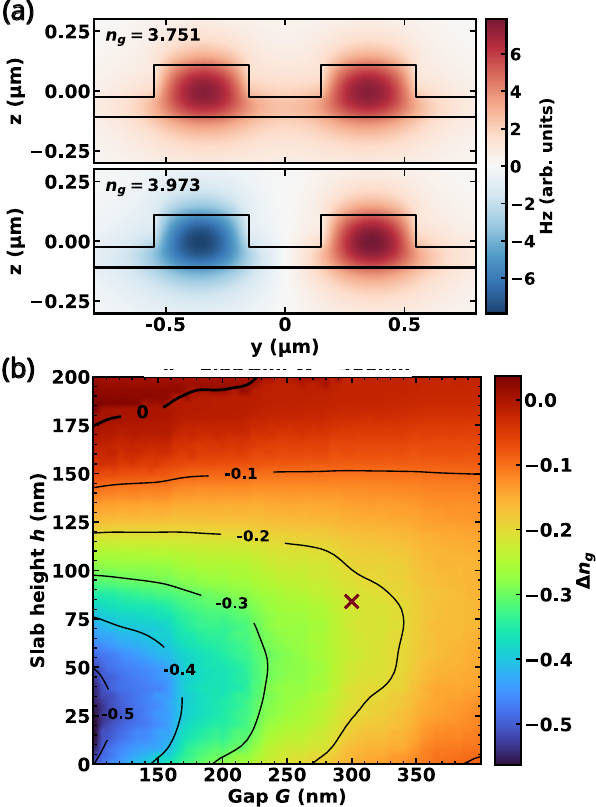}
    \caption{
    (a) MPB simulations of the two TE supermodes of the bidirectional coupler for the experimentally chosen parameters
    $(G,h,W)=(300,85,400)$~nm.
    The symmetric (top) and antisymmetric (bottom) supermodes are shown through the $H_z$ field component (arbitrary units).
    (b) Simulated differential group index $\Delta n_g(G,h)$ for fixed width $W=400$~nm.
    The thick black contour indicates $\Delta n_g=0$, corresponding to wavelength-independent splitting.
    The red cross marks the experimentally selected parameters.}
    \label{fig:coupler_design_MPB}
\end{figure}

Figure~\ref{fig:coupler_design_MPB}(a) shows the two TE supermodes computed with the Python MIT photonic bands (MPB) library for the experimentally selected geometry $(G,h,W)=(300,85,400)$~nm.

Figure~\ref{fig:coupler_design_MPB}(b) presents the simulated map of the differential group index $\Delta n_g$ as a function of gap $G$ and slab height $h$, for fixed width $W=400$~nm.
The thick black contour corresponds to $\Delta n_g=0$, which defines the condition for first-order wavelength-independent coupling. The experimentally chosen parameters (red cross) represent a compromise between (i) a small $|\Delta n_g|$ to reduce wavelength dependence, (ii) a compact coupling length $l_{50}$ of the order of $10$~µm, and (iii) moderate bending losses compatible with the graph layout. They yield $\Delta n_g = -0.22$ and $l_{50} = 8.3~\mu\mathrm{m}$.\\

\textbf{Experimental characterization.} Figure~\ref{fig:coupler_design_exp} shows the experimentally measured transmissions $P_{1\rightarrow 3}$ and $P_{1\rightarrow 4}$ over a 160~nm bandwidth (1480-1640~nm).
The data are normalized by the transmission of a reference bus waveguide not connected to a network and fabricated on the same chip.

The coupler is close to 50:50 around 1500~nm, in agreement with the designed $l_{50}$.
However, a residual wavelength dependence remains, consistent with the finite value of $\Delta n_g$ obtained in simulations.
The splitting progressively deviates from 50:50 beyond 1600~nm, leading to reduced wave function mixing inside the graphs. For this reason, resonances above 1600~nm are not included in the spectral statistics analysis presented in Fig.~\ref{fig:NNSD}.

Overall, the combined MPB optimization and experimental characterization demonstrate that the selected geometry provides a sufficiently broadband and compact implementation of near-balanced vertices, compatible with the requirements of quantum-graph-based spectral analysis.

\clearpage

\begin{center}
 \Large\textbf{Supplemental Material}\end{center}
 \normalsize

\makeatletter
\let\@sectioncntformat\@undefined
\def\@hangfrom@section#1#2#3{\@hangfrom{#1#2}\MakeTextUppercase{#3}}
\setcounter{secnumdepth}{4}
\renewcommand{\p@subsection}{}
\makeatother
\setcounter{section}{0}
\setcounter{figure}{0}
\setcounter{equation}{0}
\renewcommand{\thesection}{S\arabic{section}}
\renewcommand{\thesubsection}{S\arabic{section}.\arabic{subsection}}
\renewcommand{\thefigure}{S\arabic{figure}}
\renewcommand{\theequation}{S\arabic{equation}}


\section{Graph Topologies}
\label{app:graph_topo}

\subsection{Connectivity and metric structure}

\begin{figure}[ht]
    \centering
    \includegraphics[width=1.0\linewidth]{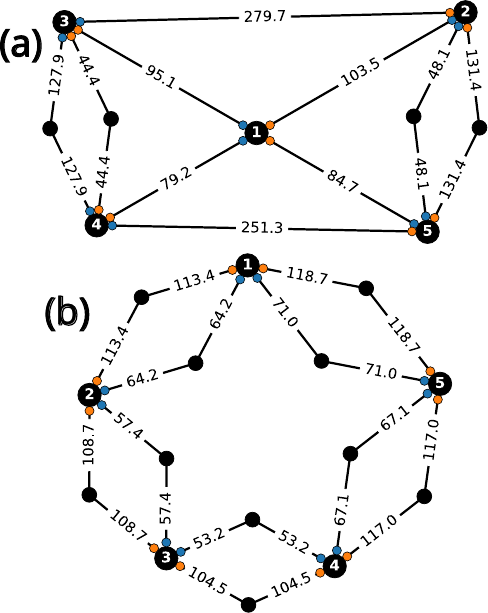}
    \caption{
        Schematic representations of (a) the BTG and (b) the FG. Numbered vertices correspond to actual bidirectional couplers, while unnumbered black dots
        represent auxiliary vertices introduced for numerical implementation.
        The incommensurate bond lengths \(L_b\) (in micrometers) are indicated on each segment.
        Each bidirectional coupler is represented by two pairs of colored ports (blue and orange): light
        injected into a port of one color exits the coupler through the ports of the other color.
    }
    \label{fig:topo}
\end{figure}

We investigate two five-vertex photonic graphs: the bow-tie graph (BTG) and the flower graph (FG), shown in Fig.~\ref{fig:topo}.
In both cases, vertices correspond to balanced $2:2$ bidirectional couplers, and bonds correspond to single-mode silicon waveguides of incommensurate lengths. This incommensurability suppresses trivial length degeneracies and ensures a well-defined semiclassical limit.

The total geometric length is
\begin{equation}
    L = \sum_b L_b,
\end{equation}
which yields $L_{\mathrm{BTG}} = 1597.1~\text{µm}$, and $L_{\mathrm{FG}}  = 1750.4~\text{µm}$.

Both graphs contain multiple bonds connecting the same pairs of vertices.
While such parallel connections are not optimal for maximizing topological complexity in the strict graph-theoretic sense, they constitute an important experimental compromise: they enhance connectivity without introducing physical waveguide crossings, which would considerably complicate fabrication and introduce uncontrolled losses.
In particular, the BTG was designed as a trade-off between strong connectivity (favoring classical mixing) and a topology that can be realized in a plane without crossings.

\subsection{Perron--Frobenius operator and classical dynamics}
\label{app:PF}

\begin{figure}[ht]
    \centering
    \includegraphics[width=1.0\linewidth]{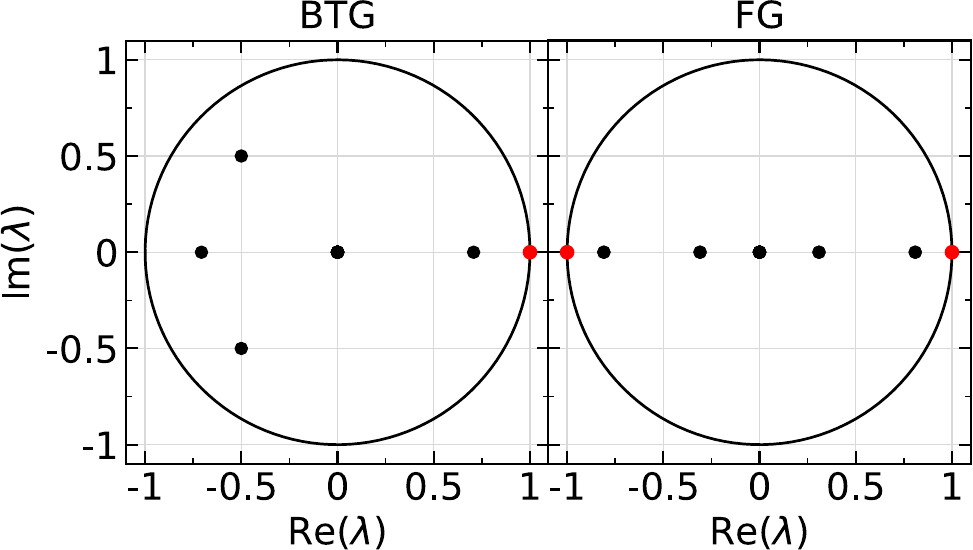}
    \caption{
        Spectra of the Perron--Frobenius operator $\mathcal{F}$ for the BTG (left) and FG (right).
        Eigenvalues are shown in the complex plane together with the unit circle.
        Black dots denote eigenvalues strictly inside the unit disk, while red dots correspond
        to eigenvalues on the unit circle. Some eigenvalues are degenerate.
        The BTG exhibits a single eigenvalue on the unit circle ($\lambda=1$), indicating
        irreducibility and aperiodicity (mixing).
        The FG possesses an additional eigenvalue at $\lambda=-1$, implying periodic dynamics and the absence of mixing despite ergodicity.
    }
    \label{fig:PFspectra}
\end{figure}

In the ray (classical) limit, propagation on the graph is described by a discrete-time Markov process on directed bonds.
Following the arguments of Ref.~\cite{gnutzmann2006}, one can deduce whether a graph is ergodic and/or mixing from the spectrum of the Perron--Frobenius-operator defined in Eq.~\eqref{eq:def-Perron-Frobenus} of the main text. Its is shown for BTG and FG in Fig.~\ref{fig:PFspectra}.

\textbf{Bow-tie graph.} The spectrum of the BTG has a unique eigenvalue on the unit circle ($\lambda=1$). All other eigenvalues satisfy $|\lambda|<1$, implying that the classical dynamics is mixing \cite{gnutzmann2006}. The absolute spectral gap~\cite{berkolaiko2026} is defined by
\begin{equation}
    \Delta = 1 - \max_{\lambda \neq 1} |\lambda|,
\end{equation}
which yields $\Delta_{\mathrm{BTG}} \approx 0.29$.
This finite gap ensures exponential decay of classical correlations and places the BTG in the standard universality class for wave graphs with time-reversal symmetry.\\

\textbf{Flower graph.}
The FG spectrum possesses an additional eigenvalue at $\lambda=-1$ on the unit circle.
The presence of this second unimodular eigenvalue implies periodic (i.e., non-aperiodic) classical dynamics.
Consequently, the graph is not mixing and no spectral gap can be defined in the strict sense.
This structural difference between FG and BTG is expected to leave signatures in spectral correlations as show in Fig.~\ref{fig:NNSD} of the main text.


\section{Model for Open Graphs}
\label{app:open_graphs}

For measurements, the closed network must be coupled to an external (bus) waveguide.
This is achieved by attaching semi-infinite leads (ports) to a selected vertex of the graph, thereby turning the system into an open scattering network. Using the port numbering of  Fig.~\ref{fig:exp_graphs}(c) of the main text, the network is connected to ports 2 and 4 of the input-output coupler, while the external leads are connected to its ports 1 and 3.

As in the closed case, the field on each internal bond is written as a superposition of counter-propagating waves, and vertex scattering is described by local scattering matrices $\sigma$.
The only modification is that the input-output coupler connects internal bonds $b$ to external leads $\beta$ ($\beta=1$ or 2).
Following the standard formalism of open wave graphs~\cite{texier2001, barra2001, kottos2003}, the transmission matrix $\mathcal{T}$ is a 2$\times$2 matrix and relates the incoming amplitude at one external lead to the outgoing amplitude at the other one,
\begin{equation}
    \mathbf{a}^{\mathrm{(out)}}_{\boldsymbol{\beta}}
    = \mathcal{T}\, \mathbf{a}^{\mathrm{(in)}}_{\boldsymbol{\beta}},
\end{equation}

Similar to the closed-graph operator, the internal operator of the open graph $\tilde{\mathcal{U}}_B$ reads
\begin{equation}
    \tilde{\mathcal{U}}_B(k) = P(k)\, \tilde{\Sigma}
\end{equation}
where $\tilde{\Sigma}$ takes into account the input-output coupler.
Now, the matrix $\tilde{\mathcal{U}}_B(k)$ is \emph{subunitary}, since part of the incoming signal escapes through the external leads.
Consequently, eigenvalues $k_p$ are not obtained from a quantization condition but from the positions of the complex poles of the scattering matrix, which are determined from
\begin{equation}
    \det\!\big[\mathbb{I}-\tilde{\mathcal{U}}_B(k_p)\big]=0,
    \qquad \Imag(k_p)<0.
\end{equation}
Each resonance position can be written as
\begin{equation}
    k_p = \textrm{Re}(k_p) - i\,\frac{\Gamma_p}{2},
\end{equation}
where $\Gamma_p$ quantifies the decay rate induced by leakage into the leads.
The imaginary part therefore determines the finite lifetime of the corresponding quasi-bound state.\\

\begin{figure}[ht]
    \centering
    \includegraphics[width=0.9\linewidth]{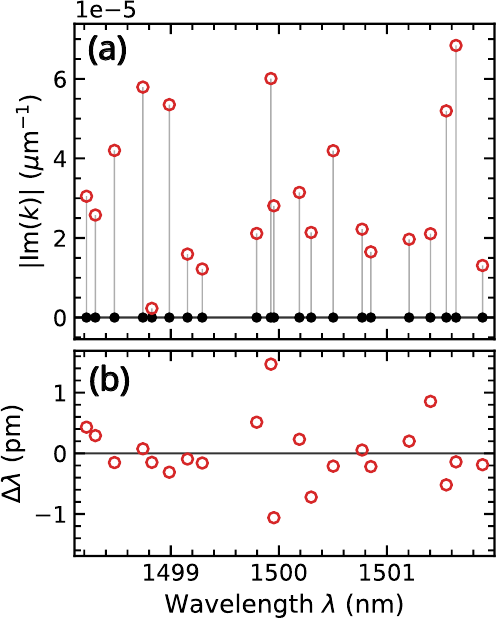}
    \caption{
    Comparison between the closed- and open-graph spectra of the BTG in the
    wavelength window 1.498--1.502~$\mu$m. (a)~Black dots denote the real
    eigennumbers obtained from the closed-graph model, while red open circles
    represent the complex poles of the open-graph scattering matrix; for each
    resonance, the vertical segment connects the closed eigenvalue to its
    corresponding open-system pole. (b)~Wavelength shift $\Delta\lambda =
    \lambda_{\mathrm{open}} - \lambda_{\mathrm{closed}}$ between each
    open-system pole (real part) and its nearest closed-model eigenvalue.
    }
    \label{fig:open_vs_closed}
\end{figure}

We use this open-graph formalism to compute the complex resonances of the BTG.
The resulting poles are compared to the real eigenvalues of the closed graph in Fig.~\ref{fig:open_vs_closed}.

As evidenced in Fig.~\ref{fig:open_vs_closed}, opening the graph produces only a small imaginary
shift while leaving the real part of the spectrum essentially unchanged: the
mean absolute wavelength deviation between open and closed resonances is
$\langle|\Delta\lambda|\rangle \approx 0.38$~pm, a relative shift of
$\approx 2.5\times10^{-7}$ of the central wavelength. Hence, in the parameter regime
investigated here, the coupling to the bus waveguide acts as a weak
perturbation of the internal dynamics. The resonance positions are therefore
governed primarily by the closed quantum map $\mathcal{U}_B(k)$, while the
open formalism mainly introduces finite lifetimes. This explains why the
closed-graph model is sufficient to capture the spectral statistics of the BTG
reported in Fig.~\ref{fig:NNSD} of the main text. The bus waveguide modifies the
linewidths but does not significantly alter the underlying level fluctuations.


\section{NNSD computation and additional spectral statistics}
\label{app:statistics}

\subsection{Weyl Law and spectral unfolding}
\label{app:unfolding}

The statistical analysis of spectral fluctuations requires separating the
smooth part of the density of states from its universal fluctuations. For
non-dispersive wave graphs (constant effective index $n_\mathrm{eff}$), the Weyl law gives the average
number of resonances below wavenumber $k$ as $\bar{N}(k)=L_\mathrm{tot}\,k/\pi$
(Sec.~5.1 of \cite{gnutzmann2006}), where $L_\mathrm{tot}=\sum_b L_b$ is the total geometric
length of the graph. In our photonic platform the propagation phase along each bond is
$n_\mathrm{eff}(k)\,k\,L_b$, with a wavenumber-dependent effective index $n_\mathrm{eff}(k)$.
Generalizing the standard derivation \cite{gnutzmann2006} to this dispersive case, the Weyl
counting function becomes
\begin{equation}
    \bar{N}(k)=\frac{n_\mathrm{eff}(k)\,L_\mathrm{tot}}{\pi}\,k + N_0,
\end{equation}
where $N_0$ is a constant offset accounting for the fact that $\bar{N}(0)$ can be non-zero, and its derivative gives the mean spectral density
\begin{equation}
    \bar{\rho}(k)=\frac{d\bar{N}}{dk}=\frac{n_g(k)\,L_\mathrm{tot}}{\pi},
\end{equation}
where the group index is defined as
\begin{equation}
    n_g(k)\equiv n_\mathrm{eff}(k)+k\,\frac{dn_\mathrm{eff}}{dk}(k).
\end{equation}

To compare spectral fluctuations with universal predictions of
random-matrix theory (RMT), the measured resonance sequence $\{k_p\}$ is
\emph{unfolded} by removing the smooth density of states using the Weyl
law,
\begin{equation}
    \tilde{k}_p = \bar{N}(k_p).
\end{equation}
This mapping produces a dimensionless spectrum with unit mean spacing,
allowing direct comparison with RMT statistics, in particular the NNSD presented in Fig.~\ref{fig:NNSD} of the main text.

\subsection{Additional Spectral Statistics}
\label{app:additional_stats}

\begin{figure}[ht]
    \centering
    \includegraphics[width=0.9\linewidth]{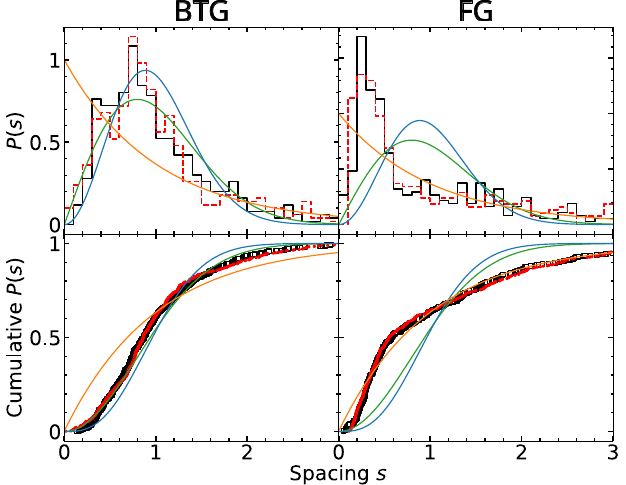}
    \caption{
        Additional short-range spectral statistics for the two photonic graphs.
        Left panels correspond to the BTG and right panels to the
    FG, using the same unfolded resonance sequences as in
        Fig.~\ref{fig:NNSD} of the main text.
        The first row reproduces the NNSD $P(s)$ of Fig.~\ref{fig:NNSD} of the main text
        for reference. The second row shows the corresponding cumulative
        distribution $I(s)=\int_0^{s} P(s')\,ds'$.
        In all panels, experimental data are shown as black continuous
        lines, while red dotted line corresponds to the prediction of the
        closed-graph model.
        The green, blue, and orange curves exhibit the corresponding results for GOE, Gaussian unitary ensemble (GUE), and Poisson statistics,
        respectively.
    }
    \label{fig:short_range}
\end{figure}

Figure~\ref{fig:short_range} presents additional short-range spectral
statistics obtained from the same unfolded resonance sequences used in
Fig.~\ref{fig:NNSD} of the main text. For both photonic graphs, the experimental data show a
good overall agreement with the numerical results for the closed-graph
model. The BTG exhibits clear level repulsion and roughly follows GOE prediction. In contrast, the FG displays
noticeable deviations from GOE statistics, with weaker level correlations
visible in the cumulative spacing distribution. It should be noted that the use of directional couplers allows to construct unidirectional graphs, for which GUE statistics is observed \cite{che2022}. However, since the BTG is bidirectional, its statistics complies with GOE statistics as expected.

\begin{figure}[ht]
    \centering
    \includegraphics[width=0.9\linewidth]{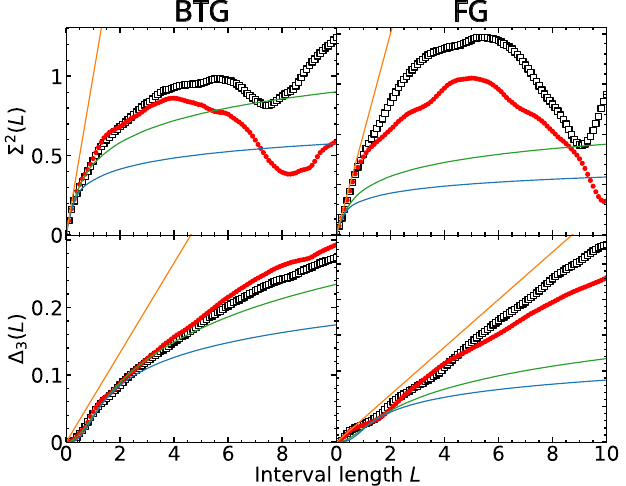}
    \caption{
        Long-range spectral correlations for the same resonance sequences as in
        Fig.~\ref{fig:NNSD} of the main text.
        Left panels correspond to the BTG and right panels to the FG.
        The first row shows the number variance $\Sigma^2(L)$, while the second
        row displays the spectral rigidity $\Delta_3(L)$.
        Experimental results are shown as black empty squares and
        numerical results for the closed-graph model as red markers.
        The green, blue, and orange curves exhibit the corresponding results for GOE, GUE, and Poisson statistics,
        respectively.
    }
    \label{fig:long_range}
\end{figure}

Long-range correlations are analyzed in Fig.~\ref{fig:long_range} through
the number variance $\Sigma^2(L)$ and the spectral rigidity $\Delta_3(L)$ \cite{bohigas1975}.
For both devices, experimental results and theoretical predictions agree
well for short intervals ($L \lesssim 3$), indicating that the model
captures the dominant spectral correlations. At larger scales, deviations
become more pronounced. Such discrepancies are expected for finite graphs
and may also arise from the presence of multiple bonds that connect the same pair of vertices, and introduce non-universal contributions to spectral statistics \cite{gnutzmann2013, dietz2024}.

Despite these deviations, the BTG retains clear signatures of GOE-like
statistics, particularly in the behavior of $\Delta_3(L)$ at short and
intermediate scales. In contrast, the FG shows weaker spectral rigidity and
a trend towards Poisson-like statistics, consistent with the reduced level
correlations already observed in the short-range analysis.

\section{Intensity distribution analysis}

\subsection{Quantifier of wave localization}
\label{app:localization}

To characterize the spatial localization of the measured optical modes,
we use two complementary quantifiers: a normalized Shannon entropy
and the inverse participation ratio (IPR).

\subsubsection{Normalized Shannon entropy}

Experimentally, the bond intensities of each wave function are extracted from the near-field
images of the resonant modes, see Fig.~\ref{fig:exp_wavefunction} of the main text.
Since the measured third-harmonic generation (THG) signal scales as
$I_{\mathrm{THG}} \propto I^3$, the fundamental intensity in each bond
is obtained by taking the cubic root of the recorded signal.

Denoting by $I_b$ the resulting intensity associated with bond $b$,
we define a discrete probability distribution over the $B$ bonds as
\begin{equation}
    p_b = \frac{I_b}{\sum_{b'=1}^{B} I_{b'}},
    \qquad
    \sum_{b=1}^{B} p_b = 1.
\end{equation}

The Shannon entropy of the mode is then
\begin{equation}
    S = - \sum_{b=1}^{B} p_b \ln p_b,
\end{equation}
and the normalized entropy is defined as
\begin{equation}
    S_{\mathrm{norm}} = \frac{S}{\ln B},
\end{equation}
such that $0 \le S_{\mathrm{norm}} \le 1$.

A value $S_{\mathrm{norm}} \approx 1$ corresponds to a mode evenly
distributed over all bonds, while $S_{\mathrm{norm}} \ll 1$ indicates
strong localization on a small subset of bonds.

Averaged over all measured modes, we obtain $S_{\mathrm{norm}} = 0.93 \pm 0.02$ experimentally, in good agreement with $S_{\mathrm{norm}} = 0.90 \pm 0.04$ inferred from the eigenfunctions of the closed-graph model.

\subsubsection{Inverse Participation Ratio (IPR)}

As an alternative and widely used localization measure, we compute the
inverse participation ratio (IPR), defined as \cite{kaplan2001}

\begin{equation}
    \mathrm{IPR} = \sum_{b=1}^{B} p_b^2.
\end{equation}

For a perfectly delocalized state with $p_b = 1/B\, \forall\, b$, the inverse participation ratio is $\mathrm{IPR} = 1/B$,
whereas for a state localized on a single bond, $\mathrm{IPR} = 1$. Thus, the IPR increases with increasing localization.

Averaged over all measured modes, we obtain $\mathrm{IPR} = 0.119 \pm 0.01$ experimentally, in good agreement with $\mathrm{IPR} = 0.139 \pm 0.02$ inferred from the eigenfunctions of the closed-graph model.


\subsection{Power budget for intensity distribution imaging}
\label{app:power_budget}

Intensity distribution imaging via third-harmonic generation (THG) requires a sufficiently high circulating power inside the resonant graph. We detail here the power budget from the fiber output to the graph interior.

\subsubsection{Fiber-to-chip injection}

An erbium-doped fiber amplifier (EDFA) delivers $P_\mathrm{fiber} = 50~\mathrm{mW}$ at the fiber output.
Fiber-to-chip coupling losses of $7.5~\mathrm{dB}$ (due to mode mismatch at the lensed-fiber/taper interface) reduce the power injected into the bus waveguide to $P_\mathrm{in} \approx 9~\mathrm{mW}$.

\subsubsection{Resonant build-up}

At resonance, the optical energy stored inside the closed graph is related to the total power loss rate of the resonator by
\begin{equation}
    U = \frac{Q_\mathrm{loaded}\, P_\mathrm{loss,total}}{\omega_0},
\end{equation}
where $\omega_0 = 2\pi c/\lambda$ and $P_\mathrm{loss,total} = (\gamma_0 + \gamma_c)\,U$ encompasses both internal losses (rate $\gamma_0$) and coupling losses to the bus waveguide (rate $\gamma_c$). At critical coupling ($\gamma_0 = \gamma_c$), both channels dissipate equal power, so $P_\mathrm{loss,total} = 2P_\mathrm{in}$, giving
\begin{equation}
    U = \frac{2\,Q_\mathrm{loaded}\, P_\mathrm{in}}{\omega_0}.
\end{equation}
This condition is approximately satisfied for the vast majority of the ${\sim}800$ resonances observed in the BTG, as evidenced by the near-complete transmission dips throughout the measured spectrum [Fig.~\ref{fig:exp_setup}(b) of the main text].

Taking $Q_\mathrm{loaded} = 1.5 \times 10^5$ and $\lambda = 1530~\mathrm{nm}$ gives $\omega_0 = 1.2 \times 10^{15}~\mathrm{rad/s}$, and therefore $U \approx 2~\mathrm{pJ}$.

The average circulating power inside the graph is then
\begin{equation}
    P_\mathrm{circ} = \frac{v_g\, U}{L} = \frac{c}{n_g} \frac{U}{L},
\end{equation}
where $v_g = c/n_g$ is the group velocity, $n_g = 3.97$ is the group index at $\lambda = 1550~\mathrm{nm}$, and $L = L_\mathrm{BTG} = 1597~\mathrm{\mu m}$ is the total geometric length of the graph. This gives $P_\mathrm{circ} \approx 100~\mathrm{mW}$, more than ten times the injected bus power, sufficient to produce a detectable THG signal.
The actual sensitivity of our setup exceeds this threshold by a large margin: Fig.~\ref{fig:exp_wavefunction}(c) of the main text shows clearly visible THG emission from the bus waveguide, whose power (${\sim}P_\mathrm{in} \approx 9$~mW) is roughly ten times lower than $P_\mathrm{circ}$. Since the THG intensity scales as the cube of the local optical power, the bus signal is approximately $10^3$ times weaker than that of the resonator yet remains detectable, attesting to the high dynamic range of the THG imaging technique.


\subsection{Experimental measurement of the effective index}
\label{app:measure_neff}

\begin{figure}[tb]
    \centering
    \includegraphics[width=\linewidth]{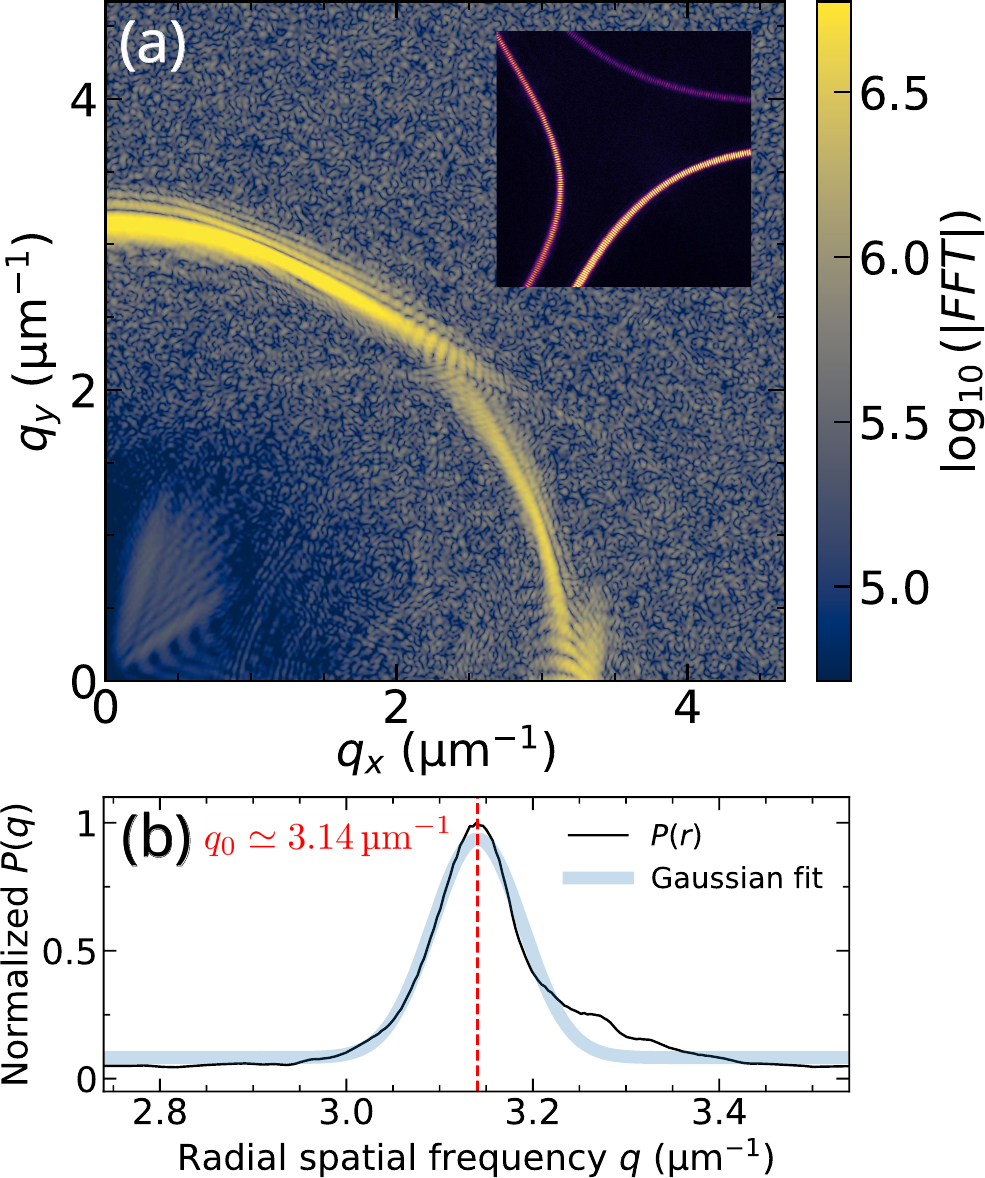}
    \caption{
    Inferring the effective index from a measured intensity distribution.
    (a) Magnitude squared of the two-dimensional Fourier transform of the experimental THG image of the guided mode at $\lambda = 1530.5$~nm in logarithmic scale. Inset: Image of the guided mode, which is also shown in Fig.~\ref{fig:exp_wavefunction}(e) of the main text.
    (b) Radial profile $P(r)$ obtained from azimuthal averaging of the Fourier amplitude (black).
    A Gaussian fit (light blue) identifies the peak position $q_0$ (dashed line), corresponding to the radius of the ring in (a).
    The effective index follows from Eq.~(\ref{eq:r0vsneff}).
    }
    \label{fig:exp_neff}
\end{figure}

The effective index of the guided mode is extracted from the spatial modulation observed in the near-field images obtained by third-harmonic generation (THG).
Because the guided field forms a standing wave along each waveguide, the measured intensity exhibits a periodic modulation whose spatial frequency is determined by the propagation constant of the mode.

To determine this spatial frequency, we compute the two-dimensional Fourier transform of the measured near-field image. An example is shown in Fig.~\ref{fig:exp_neff}(a).
In the Fourier plane, the standing-wave modulation produces an arc whose radius $q_0$ corresponds to the spatial frequency of the standing wave. This $q_0$ is obtained by azimuthal averaging of the Fourier transform.
The resulting profile, shown in Fig.~\ref{fig:exp_neff}(b), exhibits a well-defined peak at the spatial frequency $q_0$, which is determined from a Gaussian fit.

Repeating the analysis for several THG images of the same mode recorded at different locations of the photonic network yields $q_0 = \SI{3.141 \pm 0.018}{\per\micro\meter}$.
The effective index then follows from the relation
\begin{equation} \label{eq:r0vsneff}
q_0 \simeq \frac{2 n_{\mathrm{eff}}}{\lambda},
\end{equation}
which gives $n_{\mathrm{eff}} = 2.410 \pm 0.014$ at $\lambda = 1530.5$~nm in excellent agreement with MPB simulations which yields $n_\mathrm{eff} = 2.38(9)$ at $\lambda = 1531$~nm.\\

\bibliography{2025_09_05_Photonics_graphs} 

\end{document}